\documentclass[apj]{emulateapj}
\usepackage{amsmath,graphicx,epsfig,multirow,natbib,rotating,tabularx}
\usepackage[usenames]{color}
\shorttitle{ALMA Observations of SDP.9}
\shortauthors{WONG et al.}

\begin{document}

\newcommand{\comment}[1]{\textcolor{blue}{\textbf{#1}}}
\newcommand{\zl}{0.6129}
\newcommand{\zs}{1.5747}

\title{ALMA Observations of the Gravitational Lens SDP.9}
\author{
Kenneth C. Wong\altaffilmark{1,2,10},
Tsuyoshi Ishida\altaffilmark{3},
Yoichi Tamura\altaffilmark{3,4},
Sherry H. Suyu\altaffilmark{5,6,2},
Masamune Oguri\altaffilmark{7,8,9},
and
Satoki Matsushita\altaffilmark{2}
}
\altaffiltext{1}{National Astronomical Observatory of Japan, Mitaka, Tokyo 181-8588, Japan}
\altaffiltext{2}{Institute of Astronomy and Astrophysics, Academia Sinica (ASIAA), P.O. Box 23-141, Taipei 10617, Taiwan}
\altaffiltext{3}{Institute of Astronomy, University of Tokyo, 2-21-1 Osawa, Mitaka, Tokyo 181-0015, Japan}
\altaffiltext{4}{Department of Physics, Nagoya University, Furo-cho, Chikusa-ku, Nagoya 464-8601, Japan}
\altaffiltext{5}{Max-Planck-Institut f{\"u}r Astrophysik, Karl-Schwarzschild-Str.~1, 85748 Garching, Germany}
\altaffiltext{6}{Physik-Department, Technische Universit\"at M\"unchen, James-Franck-Stra\ss{}e~1, 85748 Garching, Germany}
\altaffiltext{7}{Department of Physics, The University of Tokyo, 7-3-1 Hongo, Bunkyo-ku, Tokyo 113-0033, Japan}
\altaffiltext{8}{Kavli Institute for the Physics and Mathematics of the Universe (Kavli IPMU, WPI), University of Tokyo, 5-1-5 Kashiwanoha, Kashiwa, Chiba 277-8583, Japan}
\altaffiltext{9}{Research Center for the Early Universe, University of Tokyo, Tokyo 113-0033, Japan}
\altaffiltext{10}{EACOA Fellow}

\begin{abstract}
We present long-baseline ALMA observations of the strong gravitational lens H-ATLAS J090740.0-004200 (SDP.9), which consists of an elliptical galaxy at $z_{\mathrm{L}}=0.6129$ lensing a background submillimeter galaxy into two extended arcs.  The data include Band 6 continuum observations, as well as CO $J$=6$-$5 molecular line observations, from which we measure an updated source redshift of $z_{\mathrm{S}}=1.5747$.  The image morphology in the ALMA data is different from that of the {\it HST} data, indicating a spatial offset between the stellar, gas, and dust component of the source galaxy.  We model the lens as an elliptical power law density profile with external shear using a combination of archival {\it HST} data and conjugate points identified in the ALMA data.  Our best model has an Einstein radius of $\theta_{\mathrm{E}}=0.66\pm0.01$ and a slightly steeper than isothermal mass profile slope.  We search for the central image of the lens, which can be used constrain the inner mass distribution of the lens galaxy including the central supermassive black hole, but do not detect it in the integrated CO image at a 3$\sigma$ rms level of 0.0471 Jy km s$^{-1}$.
\end{abstract}

\keywords{gravitational lensing: strong -- submillimeter: galaxies}

\section{Introduction} \label{sec:intro}
Strong gravitational lensing is a powerful tool to probe the mass structure of galaxies across a wide range of scales, from the central few hundred parsecs out to several tens of kiloparsecs.  The lensing strength depends on the total mass distribution of the lens, allowing us to study both luminous and dark matter.  Strong lensing also enables detailed studies of the background source, as the lensing magnification effect allows us to detect features in the source structure that would otherwise be unresolved.

Recent surveys at submillimeter wavelengths have revealed a population of lensed dusty star-forming galaxies at $z \sim 2-7$ \citep[e.g.,][]{negrello+2010,bussmann+2013,hezaveh+2013a,vieira+2013}.  These galaxies are among the brightest extragalactic sources at submillimeter wavelengths due to magnification bias resulting from the steepness of their luminosity function at the bright end, as well as the negative K-correction in which the rising SED from dust emission compensates dimming due to increasing cosmological distance from $z \sim 1-6$ \citep[e.g.,][]{hezavehholder2011}.  In recent years, the Atacama Large Millimeter/Submillimeter Array (ALMA) has revolutionized observational studies of star-forming galaxies as a result of its high sensitivity and spatial resolution.  By combining the power of ALMA with the natural magnification of gravitational lensing, the detailed properties of these galaxies can be studied on spatial scales of tens to hundreds of parsecs \citep[e.g.,][]{hezaveh+2013a,alma2015,bussmann+2015,dye+2015,dye+2017,hatsukade+2015,rybak+2015b}.

These systems also present a unique opportunity to study the supermassive black hole (SMBH) at the center of the lens galaxy through the brightness of the central image of the lens \citep[e.g.,][]{winn+2004,tamura+2015,wong+2015,quinn+2016}.  There are known correlations between the mass of a SMBH and physical properties of the bulge component of its host galaxy \citep[e.g.,][]{kormendyho2013}, such as luminosity, velocity dispersion, and stellar mass.  Determining the mass of a SMBH beyond the local universe requires an active galactic nucleus (AGN), which complicates measurements of the underlying host galaxy due to its extreme brightness \citep[although see][]{ding+2017a,ding+2017b}.  Lensed galaxies that are bright at submillimeter wavelengths, such as SDP.81 \citep{negrello+2010,alma2015}, subvert these challenges, as lens galaxies tend to be early-type galaxies with little to no emission in the submillimeter, making it easier to detect the central image \citep{hezaveh+2015}.

In this paper, we present long-baseline ALMA observations of the strong gravitational lens H-ATLAS J090740.0-004200 (hereafter SDP.9).  SDP.9 was first identified by \citet{negrello+2010} as part of a blind search of the brightest sources in the Science Demonstration Phase (SDP) of the {\it Herschel} Astrophysical Terahertz Large Area Survey \citep[H-ATLAS;][]{eales+2010}.  SDP.9 consists of a massive early-type galaxy at $z_{\mathrm{L}} = \zl$ \citep{bussmann+2013} lensing a background submillimeter galaxy at $z_{\mathrm{S}} = \zs$ into a bright arc and small counterimage.  The lens galaxy is in the outskirts of a $\mathrm{M_{200m}} \sim 10^{14} \mathrm{M_{\odot}}$ galaxy cluster identified by the CAMIRA algorithm \citep{oguri2014,oguri+2017} using data from the Hyper Suprime-Cam Subaru Strategic Program \citep{aihara+2017}.  These ALMA data provide an opportunity to study the detailed gas and dust properties of the source galaxy at a resolution that has only been attained for a handful of galaxies to date.

This paper is organized as follows.  We describe the ALMA data and reduction in Section~\ref{sec:data}.  We also summarize the archival {\it HST} data that are used in this analysis.  We describe our lens modeling procedure in Section~\ref{sec:lensmod}.  We present our main results in Section~\ref{sec:results} and summarize our conclusions in Section~\ref{sec:summary}.  Throughout this paper, we assume a flat $\Lambda$CDM cosmology with $\Omega_{\mathrm{m}}=0.3$, $\Omega_{\Lambda} = 0.7$, and $H_{0} = 70~\mathrm{km~s^{-1}~Mpc^{-1}}$.  All quantities are given in $h_{70}$ units.  At $z_{\mathrm{L}} = \zl$, the angular scale is $1\arcsec = 6.75~\mathrm{kpc}$.

\section{Data} \label{sec:data}

\subsection{ALMA Data} \label{subset:alma}
ALMA data were obtained during Cycle 3 (Program 2015.1.00415.S; PI: K. Wong).
The observations were carried out in 2015 November.
The Band 6 continuum was observed, as well as the CO $J$=6--5 line (hereafter CO(6--5); $\nu_\mathrm{rest}$=691.47\,GHz,
$\nu_\mathrm{obs}$=268.56\,GHz).
During the observations, 49 12\,m antennas were used, and baselines ranged
from 84.7\,m to 16.2\,km with the C36-7 configuration.
The total on-source time was 4526\,s under the stable weather conditions
with precipitable water vapors (PWVs) $\sim$0.50 mm.

For Band 6, a total bandwidth of 7.5\,GHz is available and divided into
four spectral windows with a width and resolution of 1.875\,GHz
and 31.2\,MHz in the time division mode (TDM), respectively.
The spectral windows are centered at 251.5, 254.0, 266.2, and 268.3\,GHz.
The last one is tuned for the CO(6--5) at the source redshift.
For calibrations, J0909+0121 and J0854+2006 are observed
as a phase calibrator and flux and phase calibrators, respectively\footnote{\hspace{-0.1em}https://almascience.nrao.edu/alma-data/calibrator-catalogue}.
The calibration is done in a standard manner and the
uncertainty of the flux calibration is $\sim$10\,\%.
We note that $\sim$50\,\% of the data are flagged because a number of antennas had a technical problem that
caused amplitude scattering for the flux and bandpass calibrators.

\begin{figure*}
\centering
\plotone{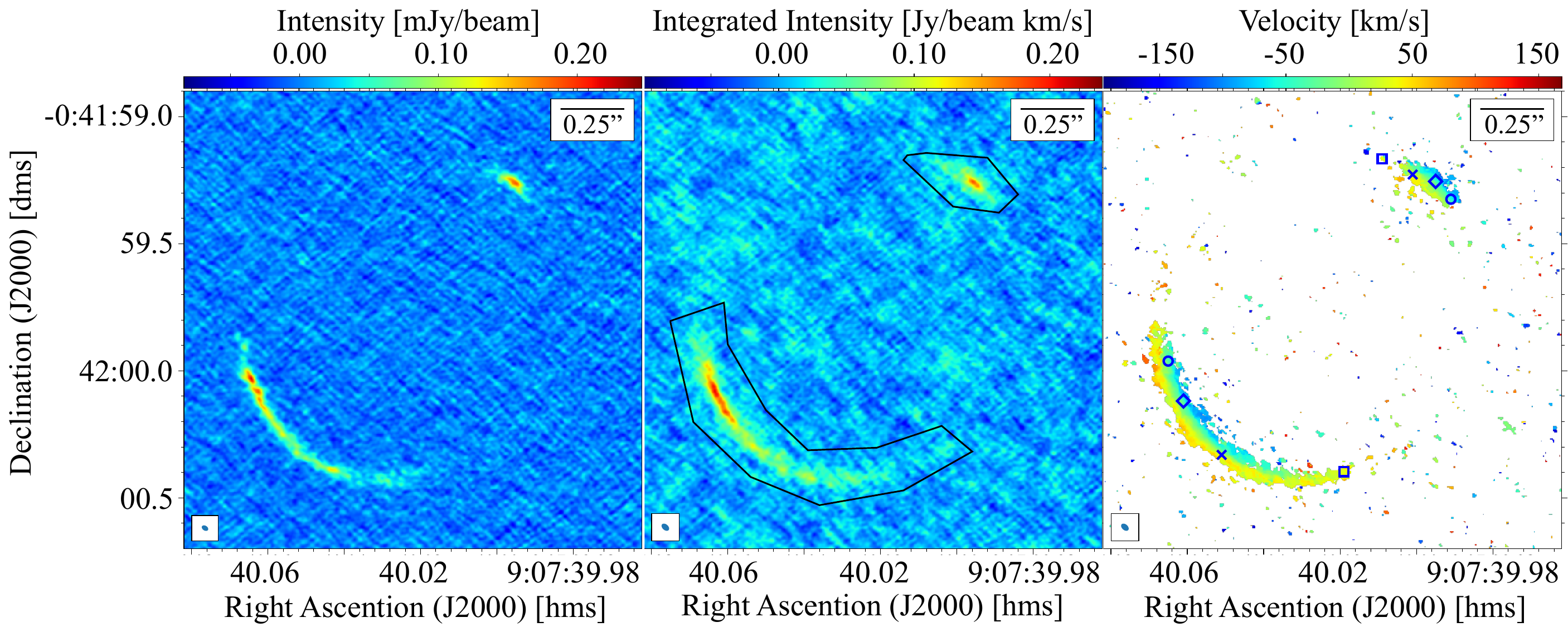}
\caption{
{\bf Left:} ALMA Band 6 continuum image with Briggs weighting (robust=0.5) and no tapering.
{\bf Center:} Velocity-integrated CO(6--5) intensity map integrated from 183190\,km\,s$^{-1}$ to 183490\,km\,s$^{-1}$ with no clipping. The black lines indicate the regions used to measure the integrated flux of the Northern and Southern arcs.
{\bf Right:} CO(6--5) velocity map with 3$\sigma$ clipping.  Small symbols represent the conjugate points used for our mass modeling, with symbols of the same shape representing image pairs.
The angular scale is indicated by the scale bar at the top right of each panel.  The beam size for each dataset is shown at the bottom left of the corresponding panel.
\label{fig:alma_data}}
\end{figure*}

We image the calibrated visibility data
using Common Astronomy Software Applications \citep[CASA;][]{mcmullin+2007} version 4.5.0.
The resulting images are shown in Figure~\ref{fig:alma_data}.
We use the CASA task \texttt{uvcontsub} to subtract the line emission
from the continuum. In executing the CASA task \texttt{clean},
we apply the natural weighting for the CO line
because we aim to maximize the sensitivity to a central image
(which is expected to be a point source).
The full width at half maximum (FWHM) of the synthesized beam 
is $35 \times 23$\,{mas} with a position angle (PA) of $49^{\circ}$ (measured counterclockwise from North),
with a root mean square (rms) noise level of 0.13\,mJy beam$^{-1}$ per 10\,km\,s$^{-1}$ channel.
For the continuum, we use both the natural and Briggs weighting (robust=0.5).
The resulting synthesized beam FWHM are $36 \times 24$\,{mas}
(PA=$50^{\circ}$) and $28 \times 19$\,{mas}
(PA=$58^{\circ}$), respectively, with rms noise levels of 0.012\,mJy beam$^{-1}$ for both.
We use the \texttt{clean} task to deconvolve the synthesized beam down to 1.5$\sigma$. We do not taper either image.

\def\arraystretch{1.5}
\begin{table}
\caption{Continuum Parameters for SDP.9 \label{tab:cont}}
\begin{ruledtabular}
\begin{tabular}{cccc}
Component &
$I_\nu^\mathrm{peak}$\tablenotemark{a} &
$F_\nu$\tablenotemark{b} &
$\theta_\mathrm{beam}$; PA\tablenotemark{c} \\
 & ($\mu$Jy beam$^{-1}$) & (mJy) & (mas; deg) \\
\tableline
Northern Arc     & $260 \pm 12$ & $1.7 \pm 0.096$ & $\ldots$ \\
Southern Arc     & $330 \pm 12$ & $8.3 \pm 0.21$  & $\ldots$ \\ 
Total & $\ldots$     & $10.0 \pm 0.23$ & $36 \times 24;\ 50$ \\
\hline
Northern Arc     & $180 \pm 12$ & $1.6 \pm 0.13$  & $\ldots$ \\
Southern Arc     & $220 \pm 12$ & $5.4 \pm 0.27$  & $\ldots$ \\
Total & $\ldots$     & $7.0 \pm 0.30$  & $28 \times 19;\ 58$ \\
\end{tabular}
\end{ruledtabular}
 \tablecomments{Values for the 260\,GHz continuum map with the natural
 weighting (first three rows) and Briggs (robust=0.5) weighting (last three rows).}
 \tablenotetext{1}{Peak intensity: The uncertainty is the root mean square
 (rms) noise level calculated from an emission-free region.}
 \tablenotetext{2}{Flux density: The intensity integrated over all spatial
 components. The uncertainty is derived as $\mathrm{rms} \times
 \sqrt{\mathrm{N_{beams}}}$, where $\mathrm{N_{beams}}$ is the number of independent synthesized beam areas included in the region.}
 \tablenotetext{3}{Beam size: Beam FWHM with PA measured counterclockwise from North.}
\end{table}

\begin{table}
\caption{CO Line Parameters for SDP.9 \label{tab:line}}
\begin{ruledtabular}
\begin{tabular}{cccc}
Component&
$I_\nu^\mathrm{peak}$\tablenotemark{a} &
$F_\nu \Delta v$\tablenotemark{b} &
$\theta_\mathrm{beam}; \mathrm{PA}$\tablenotemark{c} \\
 & (mJy beam$^{-1}$) & (Jy\,km s$^{-1}$) & (mas; deg) \\
\tableline
Northern Arc     & $1.6 \pm 0.13$ & $1.8 \pm 0.14$ & $\ldots$ \\
Southern Arc     & $1.6 \pm 0.13$ & $7.3 \pm 0.29$ & $\ldots$ \\
Total & $\ldots$       & $9.1 \pm 0.32$ & $35 \times 23;\ 49$ \\
\end{tabular}
\end{ruledtabular}
 \tablecomments{Values for the 268\,GHz CO map with the natural
 weighting.}
 \tablenotetext{1}{Peak intensity: The uncertainty is the rms
 noise level per channel (10\,km s$^{-1}$ binning) calculated from an emission-free region.}
 \tablenotetext{2}{Velocity integrated flux density:
 The intensity integrated over all spatial components and channels.
 The uncertainty is derived as $\mathrm{rms} \times
 \sqrt{\mathrm{N_{beams}}}$, where $\mathrm{N_{beams}}$ is the number of independent synthesized beam areas included in the region.}
 \tablenotetext{3}{Beam size: Beam FWHM with PA measured counterclockwise from North.}
\end{table}
\def\arraystretch{1.0}

In Figure~\ref{fig:alma_data}, we find that the source is split into two tangentially-elongated arcs, as is suggested by previous SMA observations \citep{bussmann+2013}.
The Northern arc is less extended than the Southern arc, and
some clumpy structures are visible in both arcs.
We measure the peak intensity and total flux density of each arc in both the continuum and CO maps.  The results are shown in Tables~\ref{tab:cont} and~\ref{tab:line}.
The total flux density of the continuum with the natural weighting
is consistent with previous observations conducted
with MAMBO at 1200\,$\mu$m ($F_{1200} = 7.6 \pm 1.4$\,mJy) \citep{negrello+2014},
suggesting that the missing flux is negligible.  Taking the CO line flux integrated across both arcs in each velocity bin, we measure a more accurate source redshift of $z_{\mathrm{S}} = \zs \pm 0.0002$ (95\% confidence interval) in comparison with the previous measurement of $z_{\mathrm{S}} = 1.577 \pm 0.008$ from \citet{negrello+2010}, which was determined using lower-resolution data from Z-Spec \citep{naylor+2003} on the Caltech Submillimeter Observatory.

We find no clear emission from the central demagnified image in the velocity-integrated CO(6--5) map down to a 3$\sigma$ rms level of 0.0471 Jy km s$^{-1}$.  Performing this search in the CO map rather than the continuum is important to ensure that there is no flux from low-level AGN activity in the lens galaxy that could be mistaken for a central image \citep[e.g.,][]{mckean+2005,more+2008}, although we do not detect any significant flux from the central region in the continuum data either.

\subsection{HST Data} \label{subsec:hst}
\begin{figure*}
\centering
\plotone{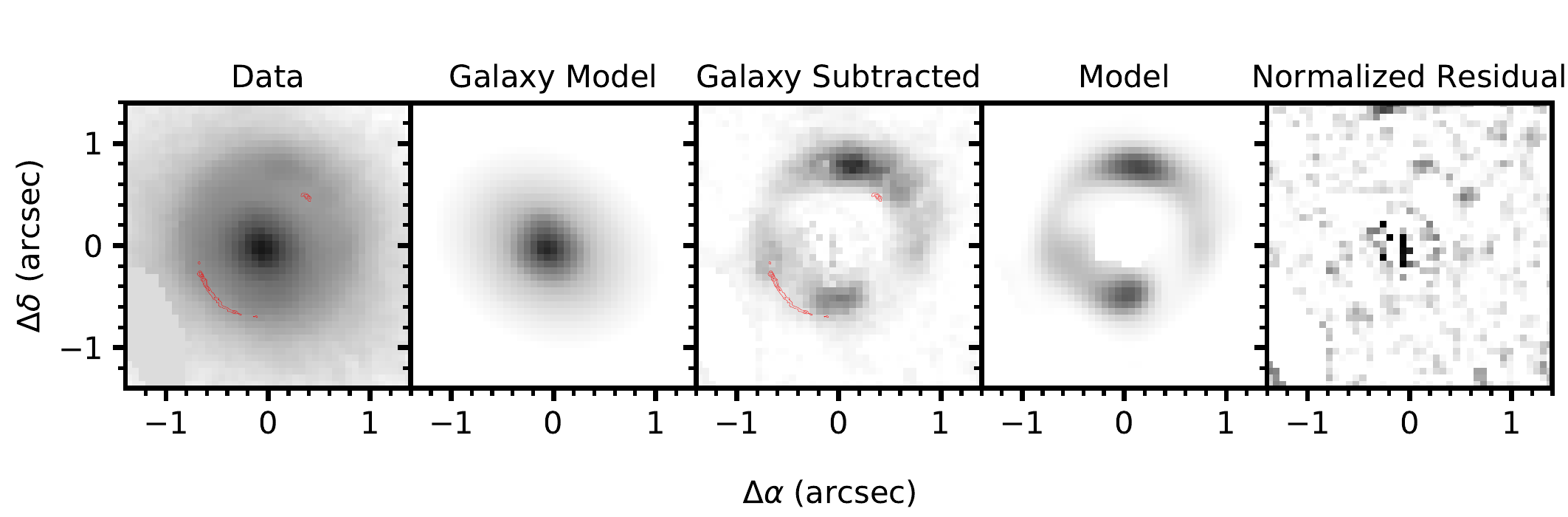}
\caption{
Image of the {\it HST} data and model reconstruction.  The panels show (from left to right) the {\it HST}/WFC3 F160W data, the best-fit lens galaxy light model, the data with the lens galaxy light model subtracted, the reconstructed image from the best-fit lens model, and the residual image normalized by the noise per pixel.  The ALMA continuum data (Figure~\ref{fig:alma_data}) are overlaid on the first and third panels (red contours) for comparison.  The configuration of the lensed images in the {\it HST} data, which probes the rest-frame optical emission from the source, is more symmetric than that of the ALMA data, which has a large Southern arc and a comparably small Northern arc, suggesting a spatial offset between the stellar component and the gas and dust component of the source.
\label{fig:sdp9_hst}}
\end{figure*}

We make use of archival {\it HST} Wide Field Camera 3 (WFC3) imaging of SDP.9.  The observations were taken in 2011 April (proposal \#12194, PI: Negrello).  These data are presented in \citet{negrello+2014} and are used for lens modeling by \citet{dye+2014}.  The observations consist of 3718 s of exposure time in the F160W filter.  The data are reduced using {\sc Drizzlepac} \citep{gonzaga+2012} with resampling to a 0\farcs065/pixel scale.  The data are shown in the left panel of Figure~\ref{fig:sdp9_hst}.

\section{Lens Model} \label{sec:lensmod}
We model the lens system using ALMA constraints and the {\it HST} images with {\sc Glee}, a lens modeling software developed by A. Halkola and S. H. Suyu \citep{suyuhalkola2010,suyu+2012}.  We have also modeled the ALMA constraints with {\sc glafic} \citep{oguri2010} and obtained similar results.  In the rest of the paper, we quote the results from {\sc Glee}.

The velocity gradient in the source galaxy is nearly orthogonal to the elongation of the lensed arcs, making a robust identification of multiple-image pair regions challenging.  We identify four pairs of conjugate points in each of the two arcs to use as lens model constraints (Figure~\ref{fig:alma_data}, right panel).  Multiple image sets are identified using the resolved velocity maps as well as iterative mass modelings with both {\sc Glee} and {\sc glafic}.  We assume a 0\farcs02 uncertainty on the positions of these points in the radial direction and 0\farcs05 uncertainty in the tangential direction.  The reduced $\chi^{2}$ of the model fit when comparing the modeled image positions to the constraints is $< 1$, suggesting that these uncertainties are overestimated, but we do not feel justified in assuming smaller uncertainties given the scatter among our attempts to identify the multiple image pairs.

We assume a cored elliptical power-law mass distribution with a central point mass (representing the SMBH) and external shear.  The use of conjugate points leads to degeneracies in the lens model, so we first fit to the extended arcs in the {\it HST} data and use the constraints on the structural parameters as priors when applying the ALMA constraints.

We first fit the lens galaxy light profile in the {\it HST} data.  The best fit comes from a combination two cored power-law profiles and a point source with their centroids linked, although we also try fitting a sum of two S\'{e}rsic profiles and a point source, which provides a comparable fit. The inner power-law profile has a core radius that is smaller than the pixel scale, which places an upper limit on it.  We then fix the galaxy light model and use the arcs to constrain the mass model.  The second panel of Figure~\ref{fig:sdp9_hst} shows our lens galaxy light model, and the third panel shows the data with the lens galaxy light subtracted.

The core radius is given a uniform prior with a maximum extent of 0\farcs065 $\approx$ 440 pc, corresponding to one pixel in the drizzled image.  The SMBH mass is given a lognormal prior in the range $7.1 
\leq \mathrm{log_{10}(M_{BH}/M_{\odot})} \leq 9.9$.  This corresponds to the 5$\sigma$ mass range expected from the relation of \citet{kormendyho2013} between SMBH mass and bulge stellar mass using the stellar mass measurement from \citet{negrello+2014}, assuming a \citet{chabrier2003} initial mass function.  Both the core radius and SMBH are unconstrained by the macro model, and thus sample the prior range.  In practice, we use importance sampling to attempt to constrain the SMBH mass from the non-detection of the central image (see Section~\ref{subsec:smbh}).

There is an offset in the relative astrometry between the {\it HST} and ALMA data, which complicates the use of the combined constraints.  We use the positions of nine stars in the {\it HST} field of view and match them to SDSS to correct for an offset in absolute astrometry, but there is still a residual scatter of $\sim 0\farcs1$.  This uncertainty, combined with an assumed $0\farcs05$ uncertainty in the absolute astrometry of SDSS and a $\sim0\farcs02$ uncertainty in the alignment of the {\it HST} and ALMA pixels, gives us a total assumed uncertainty of $\sim 0\farcs12$.  To account for this when incorporating the ALMA constraints, we give the structural parameters of the mass model a set of priors based on the posterior distributions from the fit to the {\it HST} data, then allow the centroid to vary with a Gaussian uncertainty of $0\farcs12$ when applying the ALMA  constraints.

\section{Results} \label{sec:results}

\subsection{Lens Model Results} \label{subsec:modresults}
The parameters of our best fit model are shown in Table~\ref{tab:lensmod}.  The results for the core radius and SMBH mass are not shown, as they are unconstrained by the macro model and just sample the prior (see Section~\ref{sec:lensmod}).  We also do not show the centroid coordinates due to the uncertainty in the absolute astrometry, although the model uncertainty on the centroid is $< 0.03$\arcsec.  Compared to the model of \citet{dye+2014}, we find a slightly steeper power law index and smaller Einstein radius, but the results are consistent within 2$\sigma$.  However, we find a more elliptical mass distribution ($b/a = 0.68 \pm 0.05$) compared to that of \citet{dye+2014} ($b/a = 0.88 \pm 0.06$).  The location of the lens relative to the nearby cluster would suggest an external shear in the $\theta_{\gamma} \sim -45^{\circ}$ direction.  Our lens model has a shear in the $\theta_{\gamma} = 0^{\circ}$ direction, although the shear amplitude is small ($\gamma = 0.04 \pm 0.01$) and it is not clear that the cluster should impart a large external shear at the projected distance of the lens ($\sim 80\arcsec$).  Other nearby galaxies in the field may contribute to the external shear and could have a larger influence than the cluster.  The right panel of Figure~\ref{fig:sdp9_hst} shows our best-fit model reconstruction of the {\it HST} image.

The lensed image configuration is strikingly different between the {\it HST} and ALMA data.  The {\it HST} data, which probe the rest-frame optical emission from the source, show a much more symmetric configuration, while the ALMA submillimeter data has the large bright Southern arc and the much smaller Northern arc, which is consistent with the existing SMA observations \citep{bussmann+2013}.  This difference suggests a spatial offset between the stellar component and the gas and dust component of the source galaxy, as was seen in SDP.81 \citep[e.g.,][]{dye+2015,hatsukade+2015,rybak+2015b}, as well as other lensed submillimeter galaxies \citep[e.g.,][]{iono+2006,fu+2012}.

\def\arraystretch{1.5}
\begin{table}
\caption{Lens Model Parameters \label{tab:lensmod}}
\begin{ruledtabular}
\begin{tabular}{l|c|c}
Parameter &
Description &
Posterior
\\
\tableline
$\theta_{\mathrm{E}}$ (\arcsec) &
Einstein Radius &
$0.66_{-0.01}^{+0.01}$
\\
$\gamma^{\prime}$ &
Mass Profile Slope &
$2.13_{-0.09}^{+0.10}$
\\
$b/a$ &
Minor/Major Projected Axis Ratio &
$0.68_{-0.04}^{+0.05}$
\\
$\theta$ ($^{\circ}$)\tablenotemark{a} &
Orientation &
$39_{- 3}^{+ 3}$
\\
$\gamma_{\mathrm{ext}}$ &
External Shear &
$0.04_{-0.01}^{+0.01}$
\\
$\theta_{\gamma}$ ($^{\circ}$)\tablenotemark{a}$^{,}$\tablenotemark{b} &
External Shear Orientation &
$ 0_{- 8}^{+ 8}$
\\
\end{tabular}
\end{ruledtabular}
\tablecomments{Reported values are medians, with errors corresponding to the 16th and 84th percentiles.}
\tablenotetext{1}{Angles measured East of North.}
\tablenotetext{2}{$\theta_{\gamma} = 0^{\circ}$ corresponds to shearing along North-South direction (i.e. external mass distributions East or West from the lens system).}
\end{table}
\def\arraystretch{1.0}

A more detailed model using the surface brightness distribution of the images or modeling of the ALMA visibilities \citep[e.g.,][]{rybak+2015a,rybak+2015b,hezaveh+2016,dye+2017} will provide better constraints on the mass model and a reconstruction of the intrinsic source brightness distribution (Ishida et al. in preparation), as well as potential constraints on substructure in the lens galaxy using the resolved velocity structure of the lensed images \citep[e.g.,][]{hezaveh+2013b,hezaveh+2016,inoue+2016}.

\subsection{SMBH Mass Inference} \label{subsec:smbh}
We attempt to use the non-detection of the central image to place a lower limit on the SMBH mass in a manner similar to \citet{wong+2015}, who were able to place a lower limit on the SMBH mass of SDP.81 given certain assumptions about the core radius of the lens galaxy \citep[see also][]{tamura+2015}.  We first predict the expected flux of the central image from our lens model by mapping the conjugate points used as model constraints (Section~\ref{sec:lensmod}) to the source plane.  We take the average of these mapped positions to be the location of the source in the source plane.  The mapped positions are weighted by $\sqrt{\mu_{i}}/\sigma_{i}$, where $\mu_{i}$ is the model magnification at the position of image $i$ and $\sigma_{i}$ is its associated uncertainty.  To obtain the magnification of the extended images accounting for differential magnification, we place a circular Gaussian profile at the determined source position to create a mock extended source.  The size of the Gaussian is assumed to be $\sigma = 0\farcs0168$ ($\sim140$ pc at the source redshift), which we determine by attempting to model the surface brightness distribution of the continuum image, although our results are the same even if we assume a source twice as large.  We then lens this mock source with our model and calculate the relative flux of the extended images to estimate their magnification ratio.

The non-detection likelihood of the central image in the CO map is
\begin{equation} \label{eq:lik_cen}
L_{\mathrm{flux}} = \frac{1}{\sqrt{2 \pi} \sigma_{\mathrm{CO}}} \mathrm{exp} \left[ -\frac{f\mathrm{_{cen}^{pred}}(\eta)^{2}}{2\sigma_{\mathrm{CO}}^{2}} \right],
\end{equation}
where $\sigma_{\mathrm{CO}}$ is the background rms in the velocity-integrated CO image and $f\mathrm{_{cen}^{pred}}(\eta)$ is the predicted flux density of the central image.  We determine $f\mathrm{_{cen}^{pred}}(\eta)$ by multiplying the observed integrated flux density of the Northern arc with the magnification ratio of the central image to the Northern arc.  We choose the Northern arc because of its smaller uncertainties in the differential magnification.

Despite SDP.9 being a double-image lens, which is expected to have a brighter central image than quad-image lenses in general \citep{mao+2001,keeton2003}, we are unfortunately unable to place meaningful constraints on the SMBH mass, as the predicted flux of the central image in our models is too low.  In fact, even removing the central black hole from the mass model does not produce a central image bright enough to be detectable, as the mass profile slope and core radius of the lens galaxy creates a central density that is too high.

\section{Summary} \label{sec:summary}
We present new ALMA observations of the gravitational lens system SDP.9, consisting of high-resolution long-baseline Band 6 continuum and CO $J$=6$-$5 emission line observations.  We measure an updated source redshift of $z_{\mathrm{S}} = \zs \pm 0.0002$.  The data show a distinctly different morphology than that seen in archival {\it HST} imaging, suggesting a spatial offset between the stellar distribution and the gas and dust distribution of the source galaxy.  Using a combination of {\it HST} data and conjugate points identified from the ALMA data, we model the lens system.  We search for the central image in the integrated CO image, but it is too faint to be detected down to a 3$\sigma$ rms level of 0.0471 Jy km s$^{-1}$, and we are unable to constrain the mass of the central supermassive black hole due to the high central density of the lens galaxy.  Future work will model this lens system in greater detail to study the detailed properties of the source.

\acknowledgments
This paper makes use of the following ALMA data:
   ADS/JAO.ALMA\#2015.1.00415.S. ALMA is a partnership of ESO (representing its member states),
   NSF (USA) and NINS (Japan), together with NRC (Canada), NSC and ASIAA (Taiwan), and KASI
   (Republic of Korea), in cooperation with the Republic of Chile.
   The Joint ALMA Observatory is operated by ESO, AUI/NRAO and NAOJ.
K.C.W. is supported by an EACOA Fellowship awarded by the East Asia Core
Observatories Association, which consists of the Academia Sinica
Institute of Astronomy and Astrophysics, the National Astronomical
Observatory of Japan, the National Astronomical Observatories of the
Chinese Academy of Sciences, and the Korea Astronomy and Space Science
Institute.
S.H.S. thanks the Max Planck Society for support through the Max
Planck Research Group.
M.O. is supported in part by World Premier International Research Center Initiative (WPI Initiative), MEXT, Japan, and JSPS KAKENHI Grant Number 26800093 and 15H05892.


\begin{thebibliography}{}
\expandafter\ifx\csname natexlab\endcsname\relax\def\natexlab#1{#1}\fi

\bibitem[{{Aihara} {et~al.}(2017){Aihara}, {Armstrong}, {Bickerton}, {Bosch},
  {Coupon}, {Furusawa}, {Hayashi}, {Ikeda}, {Kamata}, {Karoji}, {Kawanomoto},
  {Koike}, {Komiyama}, {Lupton}, {Mineo}, {Miyatake}, {Miyazaki}, {Morokuma},
  {Obuchi}, {Oishi}, {Okura}, {Price}, {Takata}, {Tanaka}, {Tanaka}, {Tanaka},
  {Uchida}, {Uraguchi}, {Utsumi}, {Wang}, {Yamada}, {Yamanoi}, {Yasuda},
  {Arimoto}, {Chiba}, {Finet}, {Fujimori}, {Fujimoto}, {Furusawa}, {Goto},
  {Goulding}, {Gunn}, {Harikane}, {Hattori}, {Hayashi}, {Helminiak}, {Higuchi},
  {Hikage}, {Ho}, {Hsieh}, {Huang}, {Huang}, {Imanishi}, {Iwata}, {Jaelani},
  {Jian}, {Kashikawa}, {Katayama}, {Kojima}, {Konno}, {Koshida}, {Leauthaud},
  {Lee}, {Lin}, {Lin}, {Mandelbaum}, {Matsuoka}, {Medezinski}, {Miyama},
  {Momose}, {More}, {More}, {Mukae}, {Murata}, {Murayama}, {Nagao}, {Nakata},
  {Niikura}, {Nishizawa}, {Oguri}, {Okabe}, {Ono}, {Onodera}, {Onoue}, {Ouchi},
  {Pyo}, {Shibuya}, {Shimasaku}, {Simet}, {Speagle}, {Spergel}, {Strauss},
  {Sugahara}, {Sugiyama}, {Suto}, {Suzuki}, {Tait}, {Takada}, {Terai}, {Toba},
  {Turner}, {Uchiyama}, {Umetsu}, {Urata}, {Usuda}, {Yeh}, \&
  {Yuma}}]{aihara+2017}
{Aihara}, H., {Armstrong}, R., {Bickerton}, S., {et~al.} 2017, ArXiv e-prints,
  arXiv:1702.08449

\bibitem[{{ALMA Partnership} {et~al.}(2015){ALMA Partnership}, {Vlahakis},
  {Hunter}, {Hodge}, {P{\'e}rez}, {Andreani}, {Brogan}, {Cox}, {Martin},
  {Zwaan}, {Matsushita}, {Dent}, {Impellizzeri}, {Fomalont}, {Asaki},
  {Barkats}, {Hills}, {Hirota}, {Kneissl}, {Liuzzo}, {Lucas}, {Marcelino},
  {Nakanishi}, {Phillips}, {Richards}, {Toledo}, {Aladro}, {Broguiere},
  {Cortes}, {Cortes}, {Espada}, {Galarza}, {Garcia-Appadoo}, {Guzman-Ramirez},
  {Hales}, {Humphreys}, {Jung}, {Kameno}, {Laing}, {Leon}, {Marconi},
  {Mignano}, {Nikolic}, {Nyman}, {Radiszcz}, {Remijan}, {Rod{\'o}n}, {Sawada},
  {Takahashi}, {Tilanus}, {Vila Vilaro}, {Watson}, {Wiklind}, {Ao}, {Di
  Francesco}, {Hatsukade}, {Hatziminaoglou}, {Mangum}, {Matsuda}, {van Kampen},
  {Wootten}, {de Gregorio-Monsalvo}, {Dumas}, {Francke}, {Gallardo}, {Garcia},
  {Gonzalez}, {Hill}, {Iono}, {Kaminski}, {Karim}, {Krips}, {Kurono},
  {Lonsdale}, {Lopez}, {Morales}, {Plarre}, {Videla}, {Villard}, {Hibbard}, \&
  {Tatematsu}}]{alma2015}
{ALMA Partnership}, {Vlahakis}, C., {Hunter}, T.~R., {et~al.} 2015, \apjl, 808,
  L4

\bibitem[{{Bussmann} {et~al.}(2013){Bussmann}, {P{\'e}rez-Fournon}, {Amber},
  {Calanog}, {Gurwell}, {Dannerbauer}, {De Bernardis}, {Fu}, {Harris}, {Krips},
  {Lapi}, {Maiolino}, {Omont}, {Riechers}, {Wardlow}, {Baker}, {Birkinshaw},
  {Bock}, {Bourne}, {Clements}, {Cooray}, {De Zotti}, {Dunne}, {Dye}, {Eales},
  {Farrah}, {Gavazzi}, {Gonz{\'a}lez Nuevo}, {Hopwood}, {Ibar}, {Ivison},
  {Laporte}, {Maddox}, {Mart{\'{\i}}nez-Navajas}, {Michalowski}, {Negrello},
  {Oliver}, {Roseboom}, {Scott}, {Serjeant}, {Smith}, {Smith}, {Streblyanska},
  {Valiante}, {van der Werf}, {Verma}, {Vieira}, {Wang}, \&
  {Wilner}}]{bussmann+2013}
{Bussmann}, R.~S., {P{\'e}rez-Fournon}, I., {Amber}, S., {et~al.} 2013, \apj,
  779, 25

\bibitem[{{Bussmann} {et~al.}(2015){Bussmann}, {Riechers}, {Fialkov},
  {Scudder}, {Hayward}, {Cowley}, {Bock}, {Calanog}, {Chapman}, {Cooray}, {De
  Bernardis}, {Farrah}, {Fu}, {Gavazzi}, {Hopwood}, {Ivison}, {Jarvis},
  {Lacey}, {Loeb}, {Oliver}, {P{\'e}rez-Fournon}, {Rigopoulou}, {Roseboom},
  {Scott}, {Smith}, {Vieira}, {Wang}, \& {Wardlow}}]{bussmann+2015}
{Bussmann}, R.~S., {Riechers}, D., {Fialkov}, A., {et~al.} 2015, \apj, 812, 43

\bibitem[{{Chabrier}(2003)}]{chabrier2003}
{Chabrier}, G. 2003, \pasp, 115, 763

\bibitem[{{Ding} {et~al.}(2017{\natexlab{a}}){Ding}, {Liao}, {Treu}, {Suyu},
  {Chen}, {Auger}, {Marshall}, {Agnello}, {Courbin}, {Nierenberg}, {Rusu},
  {Sluse}, {Sonnenfeld}, \& {Wong}}]{ding+2017a}
{Ding}, X., {Liao}, K., {Treu}, T., {et~al.} 2017{\natexlab{a}}, \mnras, 465,
  4634

\bibitem[{{Ding} {et~al.}(2017{\natexlab{b}}){Ding}, {Treu}, {Suyu}, {Wong},
  {Morishita}, {Park}, {Sluse}, {Auger}, {Agnello}, {Bennert}, \&
  {Collett}}]{ding+2017b}
{Ding}, X., {Treu}, T., {Suyu}, S.~H., {et~al.} 2017{\natexlab{b}}, ArXiv
  e-prints, arXiv:1703.02041

\bibitem[{{Dye} {et~al.}(2014){Dye}, {Negrello}, {Hopwood}, {Nightingale},
  {Bussmann}, {Amber}, {Bourne}, {Cooray}, {Dariush}, {Dunne}, {Eales},
  {Gonzalez-Nuevo}, {Ibar}, {Ivison}, {Maddox}, {Valiante}, \&
  {Smith}}]{dye+2014}
{Dye}, S., {Negrello}, M., {Hopwood}, R., {et~al.} 2014, \mnras, 440, 2013

\bibitem[{{Dye} {et~al.}(2015){Dye}, {Furlanetto}, {Swinbank}, {Vlahakis},
  {Nightingale}, {Dunne}, {Eales}, {Smail}, {Oteo}, {Hunter}, {Negrello},
  {Dannerbauer}, {Ivison}, {Gavazzi}, {Cooray}, \& {van der Werf}}]{dye+2015}
{Dye}, S., {Furlanetto}, C., {Swinbank}, A.~M., {et~al.} 2015, \mnras, 452,
  2258

\bibitem[{{Dye} {et~al.}(2017){Dye}, {Furlanetto}, {Dunne}, {Eales},
  {Negrello}, {Nayyeri}, {van der Werf}, {Serjeant}, {Farrah}, {Michalowski},
  {Baes}, {Marchetti}, {Cooray}, \& {Riechers}}]{dye+2017}
{Dye}, S., {Furlanetto}, C., {Dunne}, L., {et~al.} 2017, ArXiv e-prints,
  arXiv:1705.05413

\bibitem[{{Eales} {et~al.}(2010){Eales}, {Dunne}, {Clements}, {Cooray}, {De
  Zotti}, {Dye}, {Ivison}, {Jarvis}, {Lagache}, {Maddox}, {Negrello},
  {Serjeant}, {Thompson}, {Van Kampen}, {Amblard}, {Andreani}, {Baes},
  {Beelen}, {Bendo}, {Benford}, {Bertoldi}, {Bock}, {Bonfield}, {Boselli},
  {Bridge}, {Buat}, {Burgarella}, {Carlberg}, {Cava}, {Chanial}, {Charlot},
  {Christopher}, {Coles}, {Cortese}, {Dariush}, {da Cunha}, {Dalton}, {Danese},
  {Dannerbauer}, {Driver}, {Dunlop}, {Fan}, {Farrah}, {Frayer}, {Frenk},
  {Geach}, {Gardner}, {Gomez}, {Gonz{\'a}lez-Nuevo}, {Gonz{\'a}lez-Solares},
  {Griffin}, {Hardcastle}, {Hatziminaoglou}, {Herranz}, {Hughes}, {Ibar},
  {Jeong}, {Lacey}, {Lapi}, {Lawrence}, {Lee}, {Leeuw}, {Liske},
  {L{\'o}pez-Caniego}, {M{\"u}ller}, {Nandra}, {Panuzzo}, {Papageorgiou},
  {Patanchon}, {Peacock}, {Pearson}, {Phillipps}, {Pohlen}, {Popescu},
  {Rawlings}, {Rigby}, {Rigopoulou}, {Robotham}, {Rodighiero}, {Sansom},
  {Schulz}, {Scott}, {Smith}, {Sibthorpe}, {Smail}, {Stevens}, {Sutherland},
  {Takeuchi}, {Tedds}, {Temi}, {Tuffs}, {Trichas}, {Vaccari}, {Valtchanov},
  {van der Werf}, {Verma}, {Vieria}, {Vlahakis}, \& {White}}]{eales+2010}
{Eales}, S., {Dunne}, L., {Clements}, D., {et~al.} 2010, \pasp, 122, 499

\bibitem[{{Fu} {et~al.}(2012){Fu}, {Jullo}, {Cooray}, {Bussmann}, {Ivison},
  {P{\'e}rez-Fournon}, {Djorgovski}, {Scoville}, {Yan}, {Riechers}, {Aguirre},
  {Auld}, {Baes}, {Baker}, {Bradford}, {Cava}, {Clements}, {Dannerbauer},
  {Dariush}, {De Zotti}, {Dole}, {Dunne}, {Dye}, {Eales}, {Frayer}, {Gavazzi},
  {Gurwell}, {Harris}, {Herranz}, {Hopwood}, {Hoyos}, {Ibar}, {Jarvis}, {Kim},
  {Leeuw}, {Lupu}, {Maddox}, {Mart{\'{\i}}nez-Navajas}, {Micha{\l}owski},
  {Negrello}, {Omont}, {Rosenman}, {Scott}, {Serjeant}, {Smail}, {Swinbank},
  {Valiante}, {Verma}, {Vieira}, {Wardlow}, \& {van der Werf}}]{fu+2012}
{Fu}, H., {Jullo}, E., {Cooray}, A., {et~al.} 2012, \apj, 753, 134

\bibitem[{{Gonzaga} {et~al.}(2012){Gonzaga}, {Hack}, {Fruchter}, \&
  {Mack}}]{gonzaga+2012}
{Gonzaga}, S., {Hack}, W., {Fruchter}, A., \& {Mack}, J. 2012, {The DrizzlePac
  Handbook}

\bibitem[{{Hatsukade} {et~al.}(2015){Hatsukade}, {Tamura}, {Iono}, {Matsuda},
  {Hayashi}, \& {Oguri}}]{hatsukade+2015}
{Hatsukade}, B., {Tamura}, Y., {Iono}, D., {et~al.} 2015, \pasj, 67, 93

\bibitem[{{Hezaveh} {et~al.}(2013{\natexlab{a}}){Hezaveh}, {Dalal}, {Holder},
  {Kuhlen}, {Marrone}, {Murray}, \& {Vieira}}]{hezaveh+2013a}
{Hezaveh}, Y., {Dalal}, N., {Holder}, G., {et~al.} 2013{\natexlab{a}}, \apj,
  767, 9

\bibitem[{{Hezaveh} \& {Holder}(2011)}]{hezavehholder2011}
{Hezaveh}, Y.~D., \& {Holder}, G.~P. 2011, \apj, 734, 52

\bibitem[{{Hezaveh} {et~al.}(2015){Hezaveh}, {Marshall}, \&
  {Blandford}}]{hezaveh+2015}
{Hezaveh}, Y.~D., {Marshall}, P.~J., \& {Blandford}, R.~D. 2015, \apjl, 799,
  L22

\bibitem[{{Hezaveh} {et~al.}(2013{\natexlab{b}}){Hezaveh}, {Marrone},
  {Fassnacht}, {Spilker}, {Vieira}, {Aguirre}, {Aird}, {Aravena}, {Ashby},
  {Bayliss}, {Benson}, {Bleem}, {Bothwell}, {Brodwin}, {Carlstrom}, {Chang},
  {Chapman}, {Crawford}, {Crites}, {De Breuck}, {de Haan}, {Dobbs}, {Fomalont},
  {George}, {Gladders}, {Gonzalez}, {Greve}, {Halverson}, {High}, {Holder},
  {Holzapfel}, {Hoover}, {Hrubes}, {Husband}, {Hunter}, {Keisler}, {Lee},
  {Leitch}, {Lueker}, {Luong-Van}, {Malkan}, {McIntyre}, {McMahon}, {Mehl},
  {Menten}, {Meyer}, {Mocanu}, {Murphy}, {Natoli}, {Padin}, {Plagge},
  {Reichardt}, {Rest}, {Ruel}, {Ruhl}, {Sharon}, {Schaffer}, {Shaw},
  {Shirokoff}, {Stalder}, {Staniszewski}, {Stark}, {Story}, {Vanderlinde},
  {Wei{\ss}}, {Welikala}, \& {Williamson}}]{hezaveh+2013b}
{Hezaveh}, Y.~D., {Marrone}, D.~P., {Fassnacht}, C.~D., {et~al.}
  2013{\natexlab{b}}, \apj, 767, 132

\bibitem[{{Hezaveh} {et~al.}(2016){Hezaveh}, {Dalal}, {Marrone}, {Mao},
  {Morningstar}, {Wen}, {Blandford}, {Carlstrom}, {Fassnacht}, {Holder},
  {Kemball}, {Marshall}, {Murray}, {Perreault Levasseur}, {Vieira}, \&
  {Wechsler}}]{hezaveh+2016}
{Hezaveh}, Y.~D., {Dalal}, N., {Marrone}, D.~P., {et~al.} 2016, \apj, 823, 37

\bibitem[{{Inoue} {et~al.}(2016){Inoue}, {Minezaki}, {Matsushita}, \&
  {Chiba}}]{inoue+2016}
{Inoue}, K.~T., {Minezaki}, T., {Matsushita}, S., \& {Chiba}, M. 2016, \mnras,
  457, 2936

\bibitem[{{Iono} {et~al.}(2006){Iono}, {Peck}, {Pope}, {Borys}, {Scott},
  {Wilner}, {Gurwell}, {Ho}, {Yun}, {Matsushita}, {Petitpas}, {Dunlop},
  {Elvis}, {Blain}, \& {Le Floc'h}}]{iono+2006}
{Iono}, D., {Peck}, A.~B., {Pope}, A., {et~al.} 2006, \apjl, 640, L1

\bibitem[{{Keeton}(2003)}]{keeton2003}
{Keeton}, C.~R. 2003, \apj, 582, 17

\bibitem[{{Kormendy} \& {Ho}(2013)}]{kormendyho2013}
{Kormendy}, J., \& {Ho}, L.~C. 2013, \araa, 51, 511

\bibitem[{{Mao} {et~al.}(2001){Mao}, {Witt}, \& {Koopmans}}]{mao+2001}
{Mao}, S., {Witt}, H.~J., \& {Koopmans}, L.~V.~E. 2001, \mnras, 323, 301

\bibitem[{{McKean} {et~al.}(2005){McKean}, {Browne}, {Jackson}, {Koopmans},
  {Norbury}, {Treu}, {York}, {Biggs}, {Blandford}, {de Bruyn}, {Fassnacht},
  {Mao}, {Myers}, {Pearson}, {Phillips}, {Readhead}, {Rusin}, \&
  {Wilkinson}}]{mckean+2005}
{McKean}, J.~P., {Browne}, I.~W.~A., {Jackson}, N.~J., {et~al.} 2005, \mnras,
  356, 1009

\bibitem[{{McMullin} {et~al.}(2007){McMullin}, {Waters}, {Schiebel}, {Young},
  \& {Golap}}]{mcmullin+2007}
{McMullin}, J.~P., {Waters}, B., {Schiebel}, D., {Young}, W., \& {Golap}, K.
  2007, in Astronomical Society of the Pacific Conference Series, Vol. 376,
  Astronomical Data Analysis Software and Systems XVI, ed. R.~A. {Shaw},
  F.~{Hill}, \& D.~J. {Bell}, 127

\bibitem[{{More} {et~al.}(2008){More}, {McKean}, {Muxlow}, {Porcas},
  {Fassnacht}, \& {Koopmans}}]{more+2008}
{More}, A., {McKean}, J.~P., {Muxlow}, T.~W.~B., {et~al.} 2008, \mnras, 384,
  1701

\bibitem[{{Naylor} {et~al.}(2003){Naylor}, {Ade}, {Bock}, {Bradford},
  {Dragovan}, {Duband}, {Earle}, {Glenn}, {Matsuhara}, {Nguyen}, {Yun}, \&
  {Zmuidzinas}}]{naylor+2003}
{Naylor}, B.~J., {Ade}, P.~A.~R., {Bock}, J.~J., {et~al.} 2003, in \procspie,
  Vol. 4855, Millimeter and Submillimeter Detectors for Astronomy, ed. T.~G.
  {Phillips} \& J.~{Zmuidzinas}, 239--248

\bibitem[{{Negrello} {et~al.}(2010){Negrello}, {Hopwood}, {De Zotti}, {Cooray},
  {Verma}, {Bock}, {Frayer}, {Gurwell}, {Omont}, {Neri}, {Dannerbauer},
  {Leeuw}, {Barton}, {Cooke}, {Kim}, {da Cunha}, {Rodighiero}, {Cox},
  {Bonfield}, {Jarvis}, {Serjeant}, {Ivison}, {Dye}, {Aretxaga}, {Hughes},
  {Ibar}, {Bertoldi}, {Valtchanov}, {Eales}, {Dunne}, {Driver}, {Auld},
  {Buttiglione}, {Cava}, {Grady}, {Clements}, {Dariush}, {Fritz}, {Hill},
  {Hornbeck}, {Kelvin}, {Lagache}, {Lopez-Caniego}, {Gonzalez-Nuevo}, {Maddox},
  {Pascale}, {Pohlen}, {Rigby}, {Robotham}, {Simpson}, {Smith}, {Temi},
  {Thompson}, {Woodgate}, {York}, {Aguirre}, {Beelen}, {Blain}, {Baker},
  {Birkinshaw}, {Blundell}, {Bradford}, {Burgarella}, {Danese}, {Dunlop},
  {Fleuren}, {Glenn}, {Harris}, {Kamenetzky}, {Lupu}, {Maddalena}, {Madore},
  {Maloney}, {Matsuhara}, {Micha{\l}owski}, {Murphy}, {Naylor}, {Nguyen},
  {Popescu}, {Rawlings}, {Rigopoulou}, {Scott}, {Scott}, {Seibert}, {Smail},
  {Tuffs}, {Vieira}, {van der Werf}, \& {Zmuidzinas}}]{negrello+2010}
{Negrello}, M., {Hopwood}, R., {De Zotti}, G., {et~al.} 2010, Science, 330, 800

\bibitem[{{Negrello} {et~al.}(2014){Negrello}, {Hopwood}, {Dye}, {Cunha},
  {Serjeant}, {Fritz}, {Rowlands}, {Fleuren}, {Bussmann}, {Cooray},
  {Dannerbauer}, {Gonzalez-Nuevo}, {Lapi}, {Omont}, {Amber}, {Auld}, {Baes},
  {Buttiglione}, {Cava}, {Danese}, {Dariush}, {De Zotti}, {Dunne}, {Eales},
  {Ibar}, {Ivison}, {Kim}, {Leeuw}, {Maddox}, {Micha{\l}owski}, {Massardi},
  {Pascale}, {Pohlen}, {Rigby}, {Smith}, {Sutherland}, {Temi}, \&
  {Wardlow}}]{negrello+2014}
{Negrello}, M., {Hopwood}, R., {Dye}, S., {et~al.} 2014, \mnras, 440, 1999

\bibitem[{{Oguri}(2010)}]{oguri2010}
{Oguri}, M. 2010, \pasj, 62, 1017

\bibitem[{{Oguri}(2014)}]{oguri2014}
---. 2014, \mnras, 444, 147

\bibitem[{{Oguri} {et~al.}(2017){Oguri}, {Lin}, {Lin}, {Nishizawa}, {More},
  {More}, {Hsieh}, {Medezinski}, {Miyatake}, {Jian}, {Lin}, {Takada}, {Okabe},
  {Speagle}, {Coupon}, {Leauthaud}, {Lupton}, {Miyazaki}, {Price}, {Tanaka},
  {Chiu}, {Komiyama}, {Okura}, {Tanaka}, \& {Usuda}}]{oguri+2017}
{Oguri}, M., {Lin}, Y.-T., {Lin}, S.-C., {et~al.} 2017, ArXiv e-prints,
  arXiv:1701.00818

\bibitem[{{Quinn} {et~al.}(2016){Quinn}, {Jackson}, {Tagore}, {Biggs},
  {Birkinshaw}, {Chapman}, {De Zotti}, {McKean}, {P{\'e}rez-Fournon}, {Scott},
  \& {Serjeant}}]{quinn+2016}
{Quinn}, J., {Jackson}, N., {Tagore}, A., {et~al.} 2016, \mnras, 459, 2394

\bibitem[{{Rybak} {et~al.}(2015{\natexlab{a}}){Rybak}, {McKean}, {Vegetti},
  {Andreani}, \& {White}}]{rybak+2015a}
{Rybak}, M., {McKean}, J.~P., {Vegetti}, S., {Andreani}, P., \& {White},
  S.~D.~M. 2015{\natexlab{a}}, \mnras, 451, L40

\bibitem[{{Rybak} {et~al.}(2015{\natexlab{b}}){Rybak}, {Vegetti}, {McKean},
  {Andreani}, \& {White}}]{rybak+2015b}
{Rybak}, M., {Vegetti}, S., {McKean}, J.~P., {Andreani}, P., \& {White},
  S.~D.~M. 2015{\natexlab{b}}, \mnras, 453, L26

\bibitem[{{Suyu} \& {Halkola}(2010)}]{suyuhalkola2010}
{Suyu}, S.~H., \& {Halkola}, A. 2010, \aap, 524, A94

\bibitem[{{Suyu} {et~al.}(2012){Suyu}, {Hensel}, {McKean}, {Fassnacht}, {Treu},
  {Halkola}, {Norbury}, {Jackson}, {Schneider}, {Thompson}, {Auger},
  {Koopmans}, \& {Matthews}}]{suyu+2012}
{Suyu}, S.~H., {Hensel}, S.~W., {McKean}, J.~P., {et~al.} 2012, \apj, 750, 10

\bibitem[{{Tamura} {et~al.}(2015){Tamura}, {Oguri}, {Iono}, {Hatsukade},
  {Matsuda}, \& {Hayashi}}]{tamura+2015}
{Tamura}, Y., {Oguri}, M., {Iono}, D., {et~al.} 2015, \pasj, 67, 72

\bibitem[{{Vieira} {et~al.}(2013){Vieira}, {Marrone}, {Chapman}, {De Breuck},
  {Hezaveh}, {Wei{$\beta$}}, {Aguirre}, {Aird}, {Aravena}, {Ashby}, {Bayliss},
  {Benson}, {Biggs}, {Bleem}, {Bock}, {Bothwell}, {Bradford}, {Brodwin},
  {Carlstrom}, {Chang}, {Crawford}, {Crites}, {de Haan}, {Dobbs}, {Fomalont},
  {Fassnacht}, {George}, {Gladders}, {Gonzalez}, {Greve}, {Gullberg},
  {Halverson}, {High}, {Holder}, {Holzapfel}, {Hoover}, {Hrubes}, {Hunter},
  {Keisler}, {Lee}, {Leitch}, {Lueker}, {Luong-van}, {Malkan}, {McIntyre},
  {McMahon}, {Mehl}, {Menten}, {Meyer}, {Mocanu}, {Murphy}, {Natoli}, {Padin},
  {Plagge}, {Reichardt}, {Rest}, {Ruel}, {Ruhl}, {Sharon}, {Schaffer}, {Shaw},
  {Shirokoff}, {Spilker}, {Stalder}, {Staniszewski}, {Stark}, {Story},
  {Vanderlinde}, {Welikala}, \& {Williamson}}]{vieira+2013}
{Vieira}, J.~D., {Marrone}, D.~P., {Chapman}, S.~C., {et~al.} 2013, \nat, 495,
  344

\bibitem[{{Winn} {et~al.}(2004){Winn}, {Rusin}, \& {Kochanek}}]{winn+2004}
{Winn}, J.~N., {Rusin}, D., \& {Kochanek}, C.~S. 2004, \nat, 427, 613

\bibitem[{{Wong} {et~al.}(2015){Wong}, {Suyu}, \& {Matsushita}}]{wong+2015}
{Wong}, K.~C., {Suyu}, S.~H., \& {Matsushita}, S. 2015, \apj, 811, 115

\end{thebibliography}
\end{document}